\documentclass[sigconf,screen]{acmart}
\AtBeginDocument{%
  \providecommand\BibTeX{{%
    \normalfont B\kern-0.5em{\scshape i\kern-0.25em b}\kern-0.8em\TeX}}}

\usepackage{enumitem}
\usepackage{tikz,lipsum}
\usepackage[most]{tcolorbox}
\usepackage{multirow}
\usepackage{makecell}
\usepackage{microtype}

\setcopyright{acmlicensed}
\acmPrice{15.00}
\acmDOI{10.1145/3611643.3613888}
\acmYear{2023}
\copyrightyear{2023}
\acmSubmissionID{fse23industry-p78-p}
\acmISBN{979-8-4007-0327-0/23/12}
\acmConference[ESEC/FSE '23]{Proceedings of the 31st ACM Joint European Software Engineering Conference and Symposium on the Foundations of Software Engineering}{December 3--9, 2023}{San Francisco, CA, USA}
\acmBooktitle{Proceedings of the 31st ACM Joint European Software Engineering Conference and Symposium on the Foundations of Software Engineering (ESEC/FSE '23), December 3--9, 2023, San Francisco, CA, USA}
\received{2023-05-18}
\received[accepted]{2023-07-31}

\begin{document}


\title[Testing Real-World Healthcare IoT Application]{Testing Real-World Healthcare IoT Application: Experiences and Lessons Learned}


\author{Hassan Sartaj}
\orcid{0000-0001-5212-9787}
\affiliation{%
  \institution{Simula Research Laboratory}
  \city{Oslo}
  \country{Norway}
}\email{hassan@simula.no}

\author{Shaukat Ali}
\orcid{0000-0002-9979-3519}
\affiliation{%
  \institution{Simula Research Laboratory, Oslo Metropolitan University} 
  \city{Oslo}
  \country{Norway}
}\email{shaukat@simula.no}

\author{Tao Yue}
\orcid{0000-0003-3262-5577}
\affiliation{%
  \institution{Simula Research Laboratory} 
  \city{Oslo}
  \country{Norway}
}\email{tao@simula.no}

\author{Kjetil Moberg}
\orcid{0009-0002-5042-7371}
\affiliation{%
 \institution{Welfare Technologies Section, Oslo Kommune Helseetaten}
 \city{Oslo}
  \country{Norway}
}\email{kjetil.moberg@hel.oslo.kommune.no}

\renewcommand{\shortauthors}{Hassan Sartaj et al.}

\begin{abstract}

Healthcare Internet of Things (IoT) applications require rigorous testing to ensure their dependability. Such applications are typically integrated with various third-party healthcare applications and medical devices through REST APIs. This integrated network of healthcare IoT applications leads to REST APIs with complicated and interdependent structures, thus creating a major challenge for automated system-level testing. 
We report an industrial evaluation of a state-of-the-art REST APIs testing approach (RESTest) on a real-world healthcare IoT application. 
We analyze the effectiveness of RESTest's testing strategies regarding REST APIs failures, faults in the application, and REST API coverage, by experimenting with six REST APIs of 41 API endpoints of the healthcare IoT application. 
Results show that several failures are discovered in different REST APIs with $\approx$56\% coverage using RESTest. Moreover, nine potential faults are identified. Using the evidence collected from the experiments, we provide our experiences and lessons learned.  

\end{abstract}



\maketitle

\section{Introduction}

Healthcare Internet of Things (IoT) applications use a cloud-based architecture to provide a central access point to different stakeholders such as healthcare professionals, caretakers, and patients~\cite{elayan2021digital}. 
This is achieved by integrating various healthcare applications serving different purposes (e.g., pharmacies) and many medical devices (e.g., medicine dispensers) assigned to patients. 
Malfunctioning due to faults in the central healthcare application or integrated applications, or medical devices may have severe consequences.
The safety-critical nature of these applications requires automated testing at various levels to ensure their dependability. 
Particularly, system-level testing of such applications requires integrating various medical devices and third-party healthcare applications. 
The integration of different healthcare applications to create an IoT cloud is accomplished through Application Programming Interfaces (APIs) following Representational State Transfer (REST)~\cite{fielding2000architectural} architecture. 
This integration leads to a complicated structure of REST APIs which makes testing such APIs a challenging task.

This paper reports our work in the real-world context of Oslo City's healthcare department~\cite{oslocity}, where different healthcare IoT applications are developed and managed to deliver a wide range of healthcare services to residents of Oslo. 
The healthcare IoT applications are connected with various third-party healthcare applications through REST APIs for providing various healthcare services to patients at home care, for instance. 
Our primary objective is to test the REST APIs of such applications at the system level connected with the APIs of third-party applications in a production environment. 
The complicated nature of these REST APIs due to their dependency on different third-party APIs requires a sophisticated testing approach to explore these REST APIs thoroughly.

Several approaches have been proposed for testing of web applications REST APIs such as RESTest~\cite{martin2021restest}, RESTler~\cite{atlidakis2019restler}, Semanthesis~\cite{hatfield2022deriving}, RESTCT~\cite{wu2022combinatorial}, RestTestGen~\cite{corradini2022resttestgen}, and EvoMaster~\cite{arcuri2018evomaster}. 
Some empirical studies have been conducted to compare different approaches for testing REST APIs, such as~\cite{kim2022automated,martin2021black,arcuri2023building}.
Our goal is to assess one of the latest approaches suitable for a healthcare IoT application deployed in Oslo City to provide recommendations to the Oslo City health department for testing their current and future healthcare IoT applications.

For assessment, we select RESTest~\cite{martin2021restest} due to its advanced features, such as online testing in a production environment, handling inter-parameter dependencies, and generating realistic test data — the initial requirements of our real-world application. 
We analyze the effectiveness of all testing techniques implemented in RESTest, i.e., constraint-based testing (CBT), adaptive random testing (ART), random testing (RT), and fuzz testing (FT). The effectiveness is measured with failures in REST APIs, coverage of REST APIs, and faults in the healthcare IoT application. 
These metrics are commonly used in similar experiments~\cite{martin2022online, kim2022automated}.
We conducted an evaluation with six REST APIs with 41 API endpoints for different healthcare features. 
Results show that several failures are discovered by RESTest in six REST APIs under test. 
Moreover, with RESTest, we achieved $\approx$56\% overall coverage of REST APIs. 
Results also show that nine potential faults are found in five REST APIs by analyzing RESTest's generated failure and success reports. 
The testing techniques CBT, ART, and RT are more effective than FT in terms of failures and faults. Finally, we provide a set of lessons learned that we believe are valuable for industries working with similar applications and other healthcare departments of cities.

The upcoming part of the paper is structured as follows. 
Industrial context and challenges are described in Section~\ref{app-context}. 
Related works are outlined in Section~\ref{relatedworks}. 
Empirical evaluation and results are presented in Section~\ref{evaluation}. 
Experiences and lessons learned are discussed in Section~\ref{lessons}. 
Finally, the paper concludes in Section~\ref{conclusion}.

\section{Industrial Context and Challenges} \label{app-context} 
The national welfare technology program~\cite{wtsoslo} started by Oslo City's healthcare department~\cite{oslocity} facilitates residents with high-quality healthcare services. 
Oslo City collaborates with several third parties to develop IoT-based healthcare applications to achieve this. 
Our collaboration with Oslo City is carried out as a part of an innovation project with Oslo City's healthcare department. 
The overall goal is to build a generic test infrastructure to improve the quality of healthcare IoT applications through rigorous and automated testing of such applications. 
For this purpose, we have access to a healthcare IoT application (system under test) and the required medical devices/applications for creating an experimental setup.

The healthcare IoT application under test has several REST APIs to allow the integration of different applications and medical devices. 
These REST APIs have complicated structures due to their dependency on the APIs of different applications. 
For instance, a common procedure for testing the medical device settings feature of a healthcare IoT application is: a testing technique sends a request with test data to a healthcare application; the healthcare application receives and processes the request, creates and sends a request including test data to medical device APIs, receives a response from medical device APIs, and creates and returns a response to the testing technique. Similar is the case with other integrated applications, e.g., pharmacies.

The main challenge of testing such complex REST APIs is regarding the evolution of healthcare IoT applications, e.g., the continuous addition of medical devices from vendors, healthcare services, and software updates of third-party applications. 
This brings another challenge of analyzing the primary source (IoT application or integrated applications) of failures or faults. 
A fault in a healthcare IoT application typically leads to failures in its REST APIs. Such failures are determined by error codes 4XX and 5XX returned in response to the REST API calls.
Moreover, third-party applications define a limit on the number of API requests. Exceeding this limit may result in service blocking or damaging a medical device.
Lastly, the automated generation of realistic and domain-specific test data for healthcare IoT applications is challenging because it requires domain knowledge.

\section{Related Works}\label{relatedworks}
\textbf{Testing REST APIs.} 
Approaches for automated black-box testing of REST APIs are proposed in the literature, including RESTest~\cite{martin2021restest,martin2022online}, ARTE~\cite{alonso2022arte} — an extension to RESTest, RESTler~\cite{atlidakis2019restler,godefroid2020intelligent}, Semanthesis~\cite{hatfield2022deriving}, RESTCT~\cite{wu2022combinatorial}, RestTestGen~\cite{corradini2022resttestgen,corradini2022automated}, EvoMaster~\cite{arcuri2018evomaster,arcuri2019restful}, RapiTest~\cite{felicio2023rapitest}, and QuickREST~\cite{karlsson2020quickrest}. 
Our study utilizes RESTest, as it is a black-box REST APIs testing approach, which supports realistic test data generation~\cite{alonso2022arte} and has shown good results in testing online applications~\cite{martin2022online}.  
Open-source tools are also available for testing REST APIs such as APIFuzzer~\cite{apifuzzer}, Tcases~\cite{tcases}, and Dredd~\cite{dredd}. 
However, they did not outperform the research-based tools as reported in~\cite{kim2022automated}.

Many works are available targeting different types of testing of REST APIs such as regression testing~\cite{godefroid2020differential}, model-based testing~\cite{liu2022morest}, specification-based testing~\cite{ed2018automatic}, robustness testing~\cite{laranjeiro2021black}, metamorphic testing~\cite{segura2018metamorphic}, search-based test case improvement~\cite{stallenberg2021improving}, security testing~\cite{atlidakis2020checking}, and test input validation using deep learning~\cite{mirabella2021deep}. 
Some studies are also conducted analyzing REST APIs testing approaches/tools in different contexts such as in~\cite{kim2022automated},~\cite{martin2021black}, and~\cite{arcuri2023building}. 
We however focus on the healthcare IoT application in the context of Oslo City's health department with the ultimate goal of suggesting the necessary testing tools for continuous testing of their evolving healthcare IoT applications.

\textbf{Testing IoT Applications.}
Several studies target different aspects of testing IoT applications~\cite{dias2018brief} such as model-based conformance testing of IoT~\cite{ahmad2016model}, testing heterogeneity of IoT devices in loop~\cite{amalfitano2017towards}, identifying faults in devices integration with IoT application~\cite{wang2022understanding}, combinatorial testing and coverage criteria for IoT systems~\cite{hu2022ct}, combinatorial testing for IoT-based smart home applications~\cite{garn2022combinatorial}, and simulation of health monitoring activities of healthcare IoT applications~\cite{sotiriadis2014towards}. 
Our work focuses on testing REST APIs of a real-world healthcare IoT application using one of the latest REST APIs testing approaches.

\section{Empirical Evaluation}\label{evaluation}
We aim to evaluate the CBT, ART, RT, and FT testing techniques of RESTest~\cite{martin2021restest} for testing an operating healthcare IoT application in terms of REST APIs failures, faults in the application, and coverage of REST APIs by answering the following research questions (RQs).

\begin{itemize}[leftmargin=10pt]
    \item \textbf{RQ1:} \textit{How effective is RESTest in detecting failures in REST APIs?}
    This RQ aims to analyze the failure detection ability of RESTest with various testing techniques such that the best one can be suggested for Oslo City.  
    \item \textbf{RQ2:} \textit{How effective is RESTest in identifying faults in REST APIs under test?}
    Since not every failure observed in a REST API request leads to a potential fault, this RQ investigates faults from the test results, which helps study the relationship between failures and actual faults.   
    \item \textbf{RQ3:} \textit{How effective is RESTest with various testing techniques regarding the REST API coverage?}
    This RQ analyzes covered and uncovered aspects of REST APIs such that testing techniques can be devised to cover the uncovered aspects of REST APIs. 
\end{itemize}

\subsection{REST APIs for Evaluation}
The healthcare IoT application used for this evaluation consists of several types of REST APIs corresponding to different features, e.g., APIs for integrating various medical devices and third-party applications (pharmacies/hospitals), medical professionals, caretakers, authentication, and patients' medical records. 
We identify selection criteria from several sessions with Oslo City's technical team. 
We select a diverse set of APIs based on (i) primary concerning features to test for rapid release, (ii) involving users with different roles, (iii) including third-party applications/medical devices, and (iv) containing multiple HTTP methods such as GET, POST, and DELETE. 
We pick six APIs (41 endpoints) using these criteria.
This includes APIs for managing various alerts, addressing diverse mechanisms for authenticating different users and third-party applications, communicating with medical devices, handling patients' events/payment history/medication and measurements, and managing user profiles/tasks/accounts (as shown in Table~\ref{tab:apis}). 
The purpose of careful selection is to utilize good representative APIs for the evaluation.

\begin{table}
  \caption{Characteristics of APIs selected for evaluation, i.e., API label, \# of endpoints (EPs), and features}
  \label{tab:apis}
  \begin{tabular}{ccp{5.4cm}}
    \toprule
    \textbf{API Label}&\textbf{EPs}&\textbf{Features}\\
    \midrule
    Alerts & 2 & Assign, Close\\ 
    \makecell[t]{Authenti-\\cation} & 13 & Email, Password, SMS, Phone, IMEI, Token, Device, Log-in, Forgot Password (Send/Resend), Credentials, Emergency Profile\\ 
    Devices & 3 & Settings, Plan, Pharmacy\\
    Patients & 6 & Health data, Billings, Time logs, Tracking, Actions, Reimbursements  \\ 
    \makecell[t]{Measure-\\ments} & 5 & Single \& Multiple Inputs (Manual/Auto), Time series \\ 
    Users & 12 & Notes, Events, Threshold, Search methods, Invoicing operations, Subscriptions \\ 
  \bottomrule
\end{tabular}
\end{table}

\subsection{Evaluation Setup, Execution, and Metrics}
\textbf{Setup.} 
We acquired RESTest latest version (\textit{v1.2.0}) from GitHub. The tool needs API schema, test configurations for each API, and a properties file. 
We use the API schema in OpenAPI Specification (OAS) format for each REST API (see Table~\ref{tab:apis}). 
We define test configurations for each REST API following the guidelines in RESTest documentation. 
The property file specifies the experiment settings, including input/output paths, testing technique (CBT/ART/RT/FT), number of test cases, coverage, and path for test reports.  
We adjusted experiment settings considering the constraints specified by Oslo City's technical team. 
The service providers of third-party applications and medical devices allow a certain number of API requests for a specific duration. 
Exceeding this limit may block third-party application requests or damage medical devices. 
For this, we configure experiment settings to make a time-bound activity with a delay between two subsequent iterations. 
We configure the experiment setup to run for one hour with a delay of one second. 
The same configurations are used for each testing technique and five REST APIs (except medical device-specific APIs). 
Since medical devices require some time to process a request and respond, we configure a delay of three seconds for the APIs involving medical devices. 
For the technique-specific settings (e.g., CBT), we use the recommended default values for RESTest.

\textbf{Execution.} 
We executed experiments on two machines with specifications: (i) an 8-core CPU, 24 GB RAM, and macOS, and (ii) a four cores 3.6 GHz CPU, 32 GB RAM, and Windows OS.
Our evaluation does not consider machine-dependent parameters (e.g., time). Hence, using different machines does not affect results.

\textbf{Metrics.} 
To analyze the experiment results, we compute the number of failures, the number of faults, and the percentage of REST API coverage. 
We calculate the unique number of failures using the heuristics defined by Martin-Lopez et al.~\cite{martin2022online}. 
Based on these heuristics, the two responses are identical if the similarity between the two responses exceeds a defined threshold (in ~\cite{martin2022online}). 
Since automated identification of faults is an open problem~\cite{martin2022online}, we perform manual inspections of test reports generated by RESTest to identify faults. 
Specifically, we compare each test case result with the corresponding API specifications.
The coverage of REST APIs is measured using criteria defined by Martin-Lopez et al.~\cite{martin2019test}. 
Every criterion calculates a test suite's coverage of the number of API elements, including requests with test parameters and their responses.

\begin{table}[!t]
    \centering
    \noindent
    \caption{Results of each REST API w.r.t techniques showing the number of failures (4XX \& 5XX), faults causing failures (FF), faults from success (FS), and percentage coverage}
    \begin{tabular}{ccccccc}
    \toprule 
        \textbf{API} & \textbf{Tech.} & \textbf{4XX} & \textbf{5XX} & \textbf{FF} & \textbf{FS} & \textbf{\%Cov.}\\ \midrule
        \multirow{4}{*}{\rotatebox[origin=c]{90}{\makecell{Alerts \\ (TCs: 1000)}}}
         &CBT&114&0&0&0&65.0\\ 
         &ART&1&1&0&0&65.0\\ 
         &RT&121&0&0&0& 65.0\\ 
         &FT&0&1&0&0&70.0\\ \midrule 
         \multirow{4}{*}{\rotatebox[origin=c]{90}{\makecell{Authenti-\\cation \\ (TCs: 2821)}}}
         &CBT & 973& 0& 0 &2 (f$_1$, f$_2$)&58.0\\ 
         &ART&845&2&0&2 (f$_1$, f$_2$)&56.6\\ 
         &RT&962&0&0&2 (f$_1$, f$_2$)&58.0\\ 
         &FT&259&1&0&1 (f$_1$)&62.9\\ \midrule 
         \multirow{4}{*}{\rotatebox[origin=c]{90}{\makecell{Devices \\ (TCs: 840)}}}
         &CBT&557&0&0&1 (f$_3$)&51.4\\
         &ART&553&0&0&1 (f$_3$)&51.3\\
         &RT&559&0&0&1 (f$_3$)&51.3\\ 
         &FT&280&65&0&1 (f$_3$)&56.8\\ \midrule
         \multirow{4}{*}{\rotatebox[origin=c]{90}{\makecell{Patients \\ (TCs: 2142)}}}
         &CBT&1218&69&1 (f$_4$)&0&50.0\\ 
         &ART&176&1&1 (f$_4$)&0&50.0\\
         &RT&1210&63&1 (f$_4$)&0&50.0\\
         &FT&0&714&1 (f$_4$)&0&43.6\\ \midrule
         \multirow{4}{*}{\rotatebox[origin=c]{90}{\makecell{Measure-\\ments \\ (TCs: 1015)}}}
         &CBT&20&52&1 (f$_5$)&1 (f$_6$)&72.3\\
         &ART&1&10&1 (f$_5$)&1 (f$_6$)&61.5\\
         &RT&27&44&1 (f$_5$)& 1 (f$_6$)&75.4\\ 
         &FT&0&1&1 (f$_5$)& 1 (f$_6$)&55.4\\ \midrule 
         \multirow{4}{*}{\rotatebox[origin=c]{90}{\makecell{Users \\ (TCs: 2736)}}}
         &CBT&523&617&2 (f$_7$, f$_8$)&1 (f$_9$)&45.5\\
         &ART&683&458&2 (f$_7$, f$_8$)&1 (f$_9$)&46.1\\
         &RT&517&623&2 (f$_7$, f$_8$)&1 (f$_9$)&45.5\\ 
         &FT&0&189&1 (f$_7$)&0&48.7\\ 
    \bottomrule
    \end{tabular}
    \label{rqsresults}
\end{table}

\subsection{Results and Discussion}
Table~\ref{rqsresults} shows the overall results of each testing technique used for six REST APIs. The results of each RQ are discussed below. 

\subsubsection{RQ1: Failures in REST APIs}\label{RQ1}
The failures were identified in various REST APIs due to HTTP error messages, i.e., client-side (4XX) and server-side (5XX) error messages. 
For the Alerts, Authentication, and Devices APIs, the majority of failures were observed due to 4XX error messages in the case of each testing technique. 
This happens due to incorrect input values or types in the request body. 
In the case of Authentication API, we observed an error message (\textit{429 Too Many Requests}) indicating that the maximum attempts for login have been reached. 
Only a few failures due to 5XX error messages were found by ART and FT in the case of Alerts and Authentication APIs. 
For Devices APIs, some 5XX error messages were detected by FT only, whereas there is not even one 5XX error message detected by CBT, ART, and RT. 
In the case of APIs related to Patients, Measurements, and Users, both types of failures are identified by CBT, ART, and RT. 
However, failures identified using FT are only due to 5XX error messages.

\begin{center}
\begin{tcolorbox}[boxsep=2.5pt,left=3.5pt,right=3.5pt,top=2.5pt,bottom=2.5pt, colframe=gray!75!black, enhanced jigsaw,interior hidden,fuzzy halo=0.35mm with gray]
   \textbf{RQ1 Result:} A higher number of failures in all REST APIs were discovered using CBT and RT, whereas, with ART and FT, the majority of failures were detected in only three REST APIs. 
\end{tcolorbox} 
\end{center}

\subsubsection{RQ2: Faults in REST APIs}\label{RQ2}
To identify faults in REST APIs under test, we investigated both failure and success reports for this purpose. 
Failures in Alerts, Authentication, and Devices APIs did not lead to faults using all testing techniques (as shown in Table~\ref{rqsresults}). 
We found two faults (f$_1$ \& f$_2$) in Authentication API and one fault (f$_3$) in the Devices API by analyzing the success results. 
One fault in Authentication API is concerned with \textit{forgot password} feature in which incorrect email results in \textit{200 OK} response. 
Another fault is identified from the error message (\textit{403 Forbidden}) while authenticating through a medical device. 
The fault in Devices API is related to \textit{settings} feature. With incorrect values or empty values for device settings, a \textit{200 OK} status is received, including an error message from the third-party device APIs.  

In Patients API related to \textit{Tracking}, we identified one fault (f$_4$) corresponding to \textit{500 Internal Server Error} message generated by all testing techniques. 
This fault is discovered due to an incorrect Enum value for the code group, which should be handled during input validation. 
Another fault (f$_5$) is found in Measurements API due to \textit{500 Internal Server Error} message which is detected by each testing technique. 
This fault is generated due to measurement type mismatch. 
In addition to faults due to failures, we identified one fault (f$_6$) from successful responses. 
For the empty input measurements with the wrong device serial number or time, a \textit{200 OK} status containing an empty array of measurements is received. 
The two faults (f$_7$ \& f$_8$) are identified in Users' APIs related to Events and Search features due to failures. 
In Events API, an incorrect \textit{CategoryId} resulted in \textit{500 Internal Server Error} message with information containing database queries (f$_7$). 
The other fault (f$_8$) was discovered due to \textit{400 Bad Request} with a message showing that the search is disabled. 
Using an incorrect Social Security Number (SSN) in the search API, led to disabling the search with SSN. 
Another API for searching through SSN resulted in \textit{503 Service Unavailable} error message that originated from a third-party healthcare application. 
A fault (f$_9$) found in success reports indicating a \textit{200 OK} status with an empty array is obtained when searching a user by incorrect date of birth. 
Similarly, for an entry note from a user with an incorrect ID, a response with \textit{200 OK} status and an empty array is received. 
Though the APIs differ, the faults (f$_8$ \& f$_9$) are common in multiple REST APIs. 

All faults outlined above are reported to Oslo City and are currently in the official review process. 
We will get confirmation after a thorough analysis by the REST APIs development team of Oslo City's industry partner.

\begin{center}
\begin{tcolorbox}[boxsep=2.5pt,left=3.5pt,right=3.5pt,top=2.5pt,bottom=2.5pt, colframe=gray!75!black, enhanced jigsaw,interior hidden,fuzzy halo=0.35mm with gray]
   \textbf{RQ2 Result:} Nine potential faults are identified in five REST APIs due to failures and success results produced by CBT, ART, and RT. Whereas FT's results can only lead to three potential faults (out of nine). 
\end{tcolorbox} 
\end{center}

\subsubsection{RQ3: Coverage of REST APIs}\label{RQ3}
In the case of both Alerts and Authentication APIs, RESTest attained $\approx$60\% coverage. 
For APIs related to Devices, Patients, and Users, RESTest can only achieve $\approx$50\% coverage. 
Only for Measurements API, more than 70\% coverage is achieved with CBT and RT techniques, whereas ART and FT can only achieve 61\% and 55\% coverage, respectively. 
From the coverage results, it is analyzed that the coverage for a request's path, operations, and parameters is 100\%. 
However, the response coverage for status codes and types of status codes is nearly 50\%. Due to this, the overall REST APIs coverage is not 100\% in all cases.

\begin{center}
\begin{tcolorbox}[boxsep=2.5pt,left=3.5pt,right=3.5pt,top=2.5pt,bottom=2.5pt, colframe=gray!75!black, enhanced jigsaw,interior hidden,fuzzy halo=0.35mm with gray]
   \textbf{RQ3 Result:} The overall coverage of REST APIs achieved by RESTest's different testing techniques is $\approx$56\%. 
\end{tcolorbox} 
\end{center}

\subsection{Threats to Validity}
To reduce threats to \textbf{external validity}, we performed empirical evaluation using REST APIs targeting features critical for testing. 
We carefully selected a good representative set of REST APIs considering the testing requirements of Oslo City. 
The application used in our evaluation has a large scope and operates in multiple countries. 
Our results may not be generalizable to all healthcare applications; however, this problem is common in empirical studies~\cite{siegmund2015views}. 
In the future, we plan to include more APIs for large-scale evaluation. 
The threat to \textbf{internal validity} can occur due to the evaluation setup. 
To minimize the chances of this threat, we used APIs and documentation provided by Oslo City to create different APIs schema.
We also held various sessions with the technical team from Oslo City to demonstrate the setup and get their feedback. 
Moreover, we defined test configurations for RESTest considering test configurations used in different experiments of RESTest.   
For \textbf{conclusion and construct validity} threats, we reported the results of our evaluation using commonly used metrics in similar types of experiments, e.g., in~\cite{martin2022online}. 
To reduce potential personal bias errors during the manual inspection, we reported the experiment results to Oslo City for confirmation. 
Making a replication package or data available publicly is not possible due to non-disclosure agreements.

\section{Experiences and Lessons Learned}\label{lessons}

\noindent\textbf{Faults Identification.}
During the faults identification process, we realized that failures in REST APIs are not only potential indicators for faults in the application. 
Successful responses can also lead to faults. 
This is especially true for REST APIs communicating with REST APIs of integrated third-party healthcare applications. 
For example, we observed that sometimes an OK response is returned with a body containing an error message that is received from a third-party application or a medical device. 
It seems that a test case is passed when actually it is not. 
Therefore, identifying the root cause of a fault is tricky due to the integration of various healthcare applications with the IoT application under test.

\noindent\textbf{Domain and Context-based Test Data.}
Though RESTest tries to generate realistic test data, RESTest could not generate test data considering domain properties and particular testing contexts. 
For instance, generating a correct medication plan for a patient is not a straightforward task. 
Creating a medication plan requires information about the number of medicine doses, starting date, dose intake time, number of days to take medicine, and medication rolls of a medicine dispenser. 
This involves understanding domain properties related to medications and the context of a medicine dispenser. 
There is still a need for domain-specific and context-aware test data generation approaches.

\noindent\textbf{Valid JSON Object Generation.}
Healthcare IoT applications use JavaScript Object Notation (JSON) format for data interchange. 
Sending a valid JSON object in an HTTP request is a preliminary requirement for the application to process the request. 
An invalid JSON object leads to client error messages with 4XX status codes. 
Since RESTest does not guarantee the generation of valid JSON objects, we identified many failures due to client-side error messages. 
Further, type mismatching is another problem in creating a valid JSON object. 
This happens when an API requires certain properties in string formats. 
For example, if an API requires a phone number in string format and RESTest generates meaningful strings, the API returns a type mismatch error message. 
Therefore, Generating a valid JSON object is necessary for testing REST APIs of IoT applications.

\noindent\textbf{Tailoring Test Configurations.}
An important input of RESTest is a test configurations file that specifies API endpoints, test parameters, weights, type of test data generator, and expected outcome. 
Though RESTest can generate test configuration files, modifying them for a particular application context is needed to be done carefully. 
Among all types of test configurations, selecting a test data generator and designing its structure is crucial, especially for JSON objects with complex layouts. 
Moreover, defining test configurations for one release may not be reused for testing the next release. 
For each REST API property change, its corresponding test configurations must also be modified accordingly.

\noindent\textbf{Considering OAS Constraints.}
OAS schema allows defining non-nullable properties. 
In our experiments, we observed that RESTest generates test cases without considering non-nullable constraints on the properties. 
Violating such constraints results in input validation error messages with 4XX status codes. 
Thus, considering such OAS constraints in RESTest's test generation process is important.

\noindent\textbf{Smart Testing Strategy.}
The dependency among various REST API endpoints of a healthcare IoT application needs an intelligent testing strategy.
For example, the steps for assigning an alert (received from a patient) to a medical professional are: (i) get an unassigned alert, (ii) identify an appropriate person (doctor, nurse, caretaker, etc.) to assign the alert, and (iii) send a POST request to assign the alert. 
Each step has a separate REST API. Calling each REST API individually in different orderings will not lead to adequately testing the alert assigning scenario. 
Consequently, failures detected in this way will not help in finding faults. 
We also observed that APIs have optional properties in their schema. 
Sometimes RESTest sends a request without optional properties and expects a failure response. 
Moreover, the overall 56\% API coverage achieved by RESTest is low for IoT applications. 
Therefore, a smart testing strategy is needed by considering API dependencies, optional properties, and API coverage.

\noindent\textbf{Test Stubs Generation.}
Many REST APIs related to various integrated healthcare applications and medical devices support two-way communication. 
Testing such REST APIs requires initial input from integrated healthcare applications or medical devices. 
For example, a medical device can generate an \textit{alert} about a patient's medical condition. 
To test the \textit{assign alert} REST API, a medical device must have generated an \textit{alert}. 
Without having an \textit{alert} beforehand, generating test cases for \textit{assign alert} REST API will not lead to finding faults. 
This requires an approach to generate test stubs for integrated healthcare applications and medical devices.

\noindent\textbf{Optimized Tests Generation.}
Testing REST APIs of an IoT application in production and a rapid-release environment involves constraints such as a fixed time frame for test generation/execution and a limit on API requests to third-party applications. 
Generating and executing many test cases for such a scenario is not practical. 
Moreover, test cases with a high degree of random test data make it challenging to reproduce faults. 
Therefore, a testing approach that focuses on generating and executing an optimized set of test cases is needed. 

\noindent\textbf{Relevance of Lessons Learned.}
Our evaluation is valuable for Oslo City in terms of improving the quality of healthcare applications and developing a testing infrastructure in the future~\cite{sartaj2023hita}. 
Additionally, since REST APIs are commonly used in IoT-based applications~\cite{palma2022assessing}, the lessons presented are also relevant for practitioners developing other IoT applications.
Moreover, similar faults (the ones discussed in this paper) exist in various IoT applications, e.g., smart home systems~\cite{wang2022understanding}. 
Consequently, practitioners can benefit from implications and lessons while testing REST APIs of their respective IoT applications, e.g., smart homes and security systems.

\section{Conclusion}\label{conclusion}
We reported a real-world evaluation of RESTest for testing REST APIs for a healthcare IoT application to assess RESTest in identifying REST APIs failures, faults in the application, and REST APIs coverage. 
We evaluate different testing techniques of RESTest including CBT, ART, RT, and FT using six REST APIs. 
Results show that several failures are discovered in different REST APIs with $\approx$56\% overall coverage using RESTest.
Results also show that nine potential faults are found in five REST APIs from failure and success reports produced by RESTest. 
Moreover, testing techniques CBT, ART, and RT are more effective than FT regarding failures and faults.

\begin{acks}
This research work is a part of the WTT4Oslo project (No. 309175) funded by the Research Council of Norway. All the experiments reported in this paper are conducted in a laboratory setting of Simula Research Laboratory; therefore, they do not by any means reflect the quality of services Oslo City provides to its citizens. Moreover, these experiments do not reflect the quality of services various vendors provide to Oslo City.  
\end{acks}

\bibliographystyle{ACM-Reference-Format}
\bibliography{refs}

\end{document}